\documentclass[aps,twocolumn,showpacs,floatfix]{revtex4}

\usepackage{dcolumn}
\usepackage{graphicx}
\usepackage{amsfonts}
\usepackage{amsmath,bm,amssymb}

\begin{document}

\title{Electronic structure and orbital polarization of LaNiO$_3$ with
  a reduced coordination and under strain: first-principles study}

\author{Myung Joon Han} \affiliation{ Department of Physics, Northern
  Illinois University, De Kalb, Illinois 60115, USA}
  \affiliation{Advanced Photon Source, Argonne National Laboratory,
    9700 South Cass Avenue, Argonne, Illinois 60439, USA}

  \author{ Michel van Veenendaal} \affiliation{ Department of Physics,
    Northern Illinois University, De Kalb, Illinois 60115, USA}
  \affiliation{Advanced Photon Source, Argonne National Laboratory,
    9700 South Cass Avenue, Argonne, Illinois 60439, USA}

\date{\today }

\begin{abstract} First-principles density functional theory
  calculations have been performed to understand the electronic
  structure and orbital polarization of LaNiO$_3$ with a reduced
  coordination and under strain. From the slab calculation to simulate
  [001] surface, it is found that $d_{3z^2-r^2}$ orbital occupation is
  significantly enhanced relative to $d_{x^2-y^2}$ occupation owing to
  the reduced coordination along the perpendicular direction to the
  sample plane. Furthermore, the sign of the orbital polarization does not change
  under external strain. The results are discussed in comparison to the
  bulk and heterostructure cases, which sheds new light on the 
  understanding of the available experimental data.
\end{abstract}


\maketitle

\section{Introduction}
Understanding transition metal oxides is of perpetual interest and
importance in condensed matter physics and material science due to
their great scientific and technological potential \cite{MIT-RMP}.
Recent advances in layer-by-layer growth techniques of these compounds
in the form of thin film and heterostructures have created
considerable interest and enabled a further understanding of these
systems \cite{MRS}.  Exotic material phenomena that are clearly
distinctive from the `normal' phases include interface
superconductivity~\cite{Reyren}, charge~\cite{Ohtomo-1,Okamoto} and
orbital reconstruction~\cite{OrbReconst}, and room-temperature
ferromagnetism \cite{Luders}. Strain-mediated \cite{Liu} and electric
field control \cite{Scherwitzl} of the metal-insulator phase
transition are the other examples of thin-film phenomena.

One of the most interesting classes of transition-metal oxide systems
is thin film LaNiO$_3$ (LNO) and heterostructures of LNO and wide-gap
insulators such as LaAlO$_3$ (LAO).  Bulk LNO is the only member in
the {\it Re}NiO$_3$ series ({\it Re}: La, Pr, Nd,...{\it etc.}) that
remains metallic at all temperatures \cite{MIT-RMP,Toraance-1992}.  As
a result, thin film LNO and LNO-based heterostructures have been
studied under various strained conditions to control the correlation
strength and the metal-insulator phase transition
\cite{LNO_film_1,LNO_film_2,Liu-PRB-RC,Boris-Science} in an effort to
find possible high temperature superconductivity \cite{Hansmann} and
conductivity enhancement \cite{Son}. Strain affects most strongly the
orbital character of the nickelates
\cite{MJHan,MJHan-DMFT,Hansmann,Jak-arxiv,John-arxiv}. Recently,
several papers have focused on how changes in orbital polarization in
heterostructures and thin films affect the electronic structure and
thereby the macroscopic material properties
\cite{MJHan,MJHan-DMFT,Hansmann,Jak-arxiv,John-arxiv,May09,May10}.

An important issue is the relationship between strain and the orbital
polarization in various structural classes of nickelates. For low-spin
trivalent nickelates, the orbital polarization $P$ can be defined
simply as the difference of the two $e_g$ orbital occupations (without
renormalization factor)
\begin{equation}
  P=n_{x^2-y^2}-n_{3z^2-r^2}.
\label{Pdef}
\end{equation}
By definition, a positive orbital polarization indicates a higher
electron occupation in the $d_{x^2-y^2}$ orbital compared to the
$d_{3z^2-r^2}$ state. Recent x-ray linear dichroism (XLD) measurements
on nickelate superlattices and thin films have reported intriguing
behavior of the orbital polarization $P$ under strain. Beside the
transformation into a different structural phase \cite{Jak-arxiv}, the
polarization change under compressive and tensile strain has mainly
been analyzed using Madelung energies. However, the detailed changes
in electronic structure cannot be understood from such a simple
electrostatic approach \cite{Jak-arxiv,John-arxiv}.  For example, both
in film and in heterostructures of LNO, compressive strain make the
$d_{3z^2-r^2}$ orbital more occupied, as one would expect.  However,
tensile strain does not make the polarization go the other way which
is in a clear contradiction to Madelung picture.

In this paper, we investigate the electronic structure and orbital
occupation by using first-principles density functional theory (DFT)
calculations. Special attention has been paid to the surface Ni state
as its contribution might be significant in thin film and in
surface-sensitive measurements. The electronic structure and orbital
property are compared to the results of heterostructure and bulk
counterpart. Translational symmetry breaking along the direction
perpendicular to the sample and the reduced coordination at the
surface are found to strongly change the orbital polarization with
respect to the bulk.  The effects of strain are also studied. Our
results shed new light on understanding the available experimental
data.

\begin{figure}[t]
  \centering \includegraphics[width=8cm]{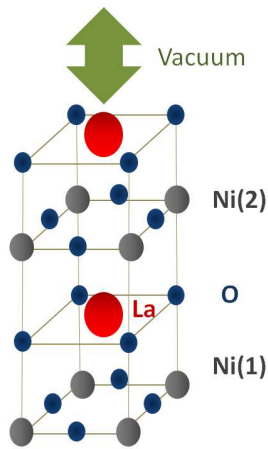}
  \caption{(Color online) Unitcell structure of the double layer slab
    calculation. Large (red), middle (gray), and small (blue) size spheres
    represent La, Ni, and O atoms, respectively. Ni(1) and Ni(2) 
    corresponds to the surface-like and bulk-like Ni sites respectively.
    \label{unitcell}}
\end{figure}

\section{Computational Details}
For the band-structure calculations, we employed Troullier-Martins
type norm-conserving pseudopotential \cite{troullier} with a partial
core correction and linear-combination-of-localized-pseudo-atomic
orbitals (LCPAO) \cite{Ozaki} as a basis set. In this pseudo-potential
generation, the semi-core 3$p$ electrons for transition metal atoms
were included as valence electrons in order to take into account the
contribution of the semi-core states to the electronic
structures. Basis orbitals are generated by a confinement potential
scheme \cite{Ozaki} with the cutoff radius 7.0 {a.u.}, 5.5 {a.u.}, and
5.0 {a.u.} for La, Ni, and O, respectively. We adopted the local
density approximation (LDA) for exchange-correlation energy functional
as parametrized by Perdew and Zunger \cite{CA}, and used energy cutoff
of 400 Ry and k-grid of $10\times 10\times 10$. All the DFT
calculations were performed using the DFT code OpenMX \cite{OpenMX}.

To simulate the [001] surface, the single and double layered slab
geometries have been investigated. The unitcell structure for the
double layer slab is presented in Fig.~\ref{unitcell}. This is thinner
than what is commonly used experimentally, but it allows us to obtain
a better understanding of the surface effects.  Vacuum layer thickness
is of 7.0~\AA ~which is large enough to make the basis function
overlap negligible.  While the polar surface may not be well
stabilized in general, these instability can be compensated by the
large covalency in the nickelates
\cite{Jak-arxiv,John-arxiv,LNO_film_1,LNO_film_2}. The geometry
relaxation has been performed with the force criterion of $10^{-3}$
Hartree/Bohr. During the relaxation process, the in-plane lattice
constant is fixed considering the substrate effect in the experimental
situation. The presence of vacuum layer naturally allows the atomic
movement along the out-of-plane axis and the volume change. We
considered the paramagnetic phase because the bulk LNO is paramagnetic
metal and there is no report on the magnetic ordering in the thin film
LNO. The orbital polarization $P$ can be calculated by integrating the
projected density-of-states (DOS) up to Fermi level.  For the atomic
charge density we used Mulliken analysis based on the generated
pseodo-atomic basis orbitals \cite{Ozaki,OpenMX}.

\begin{figure}[t]
  \centering \includegraphics[width=6cm,angle=270]{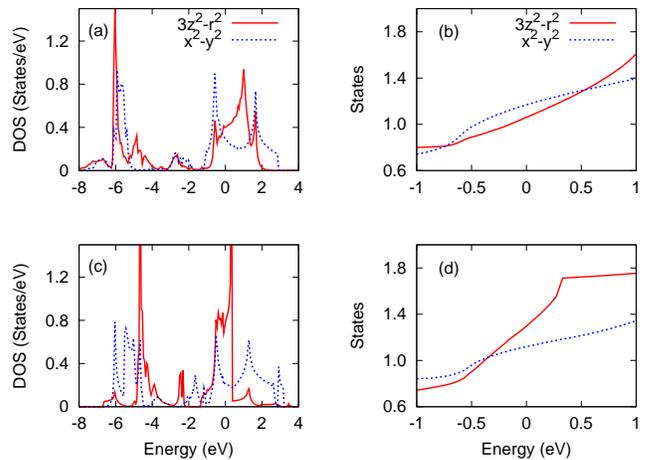}
  \caption{(Color online) (a) Projected DOS and (b) integrated DOS of
    Ni $e_g$ states in (LNO)$_1$/(LAO)$_1$ heterostructure.  (c)
    Projected DOS and (d) integrated DOS of Ni $e_g$ states in single
    layer LNO slab. Red (solid) and blue (dashed) lines correspond to
    $d_{3z^2-r^2}$ and $d_{x^2-y^2}$ states, respectively. The Fermi
    level is set to be zero.
    \label{pt_L1_N1A1}}
\end{figure}

\section{Result and Discussion}
In an ideal cubic structure of LNO, Ni$^{3+}$ ion has (formally)
singly occupied degenerate $e_g$ orbitals around the Fermi level.
This degeneracy can be lifted by, for example, making a
heterostructure with a LAO interlayer \cite{Hansmann}. It has also
been noted that the $e_g$ orbitals in LNO/LXO (X=B, Al, Ga, In)
superlattice are positively polarized without strain \cite{MJHan}. The
positive polarization is attributed to the reduced hybridization along
the $z$-direction (perpendicular to the sample plane) due to the
presence of the LAO layers.  As a result, the width of the bands with
strong $d_{3z^2-r^2}$ character is smaller compared to $d_{x^2-y^2}$.
The wider $d_{x^2-y^2}$-related bands are more occupied than the
narrower $d_{3z^2-r^2}$ band.  The calculated $e_g$ orbital DOS for
(LNO)$_1$/(LAO)$_1$ heterostructure is presented in
Fig.~\ref{pt_L1_N1A1}(a). Fig.~\ref{pt_L1_N1A1}(b), the integrated DOS
of Fig.~\ref{pt_L1_N1A1}(a), clearly shows that $d_{x^2-y^2}$ orbital
is more occupied than $d_{3z^2-r^2}$. Note that the orbital
polarization is present in the absence of strain.

\begin{figure}[t]
  \centering \includegraphics[width=6cm,angle=270]{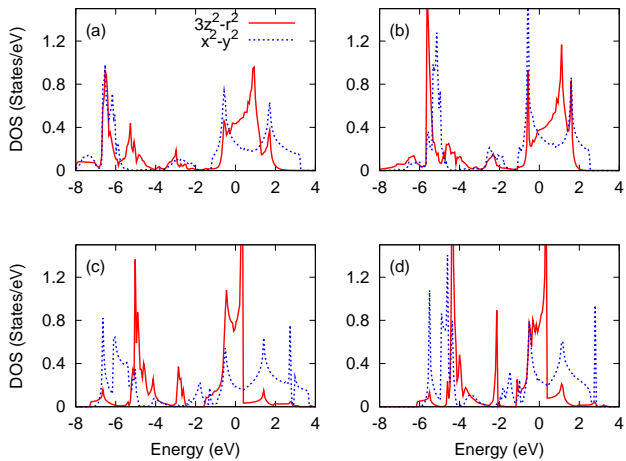}
  \caption{(Color online) Projected DOS for Ni $e_g$ states in a LNO/LAO
    heterostructure ((a) and (b)), and single layer slab ((c) and
    (d)). (a) and (c) are calculated under compressive strain  ($-3\%$
    in-plane lattice mismatch), and (b) and (d) under tensile strain
    (+3$\%$ in-plane lattice mismatch). Red (solid) and blue (dashed)
    lines correspond to $d_{3z^2-r^2}$ and $d_{x^2-y^2}$ states,
    respectively. The Fermi level is set to be zero.
    \label{pdos_L1_N1A1}}
\end{figure}

For a surface, the symmetry is also broken along the $z$-direction,
but, in addition, the Ni coordination is reduced. The calculated $e_g$
DOS of the single layer slab is shown in
Fig.~\ref{pt_L1_N1A1}(c). First, the band width of the $d_{3z^2-r^2}$
states is less than 2 eV which is not only significantly smaller than
the bulk value of $\sim 4.5$ eV, but also reduced compared to the
LNO/LAO heterostructure where the band width is larger than $\sim 2$
eV. On the other hand, the width of the $d_{x^2-y^2}$-related bands
remains comparable to the bulk value. For the heterostructure, the
narrower $d_{3z^2-r^2}$ band is located roughly in the middle of the
$d_{x^2-y^2}$ band, and the Fermi level is slightly below the middle
point of the two bands, and the $d_{x^2-y^2}$ orbital is therefore
more occupied than the $d_{3z^2-r^2}$ (Fig.~\ref{pt_L1_N1A1}(a)). As a
result, positive orbital polarization is induced as is clearly seen in
Fig.~\ref{pt_L1_N1A1}(b).  In the surface Ni site, on the other hand,
the energy level of the $d_{3z^2-r^2}$ state comes down to the bottom
of $d_{x^2-y^2}$ band due to the change in nickel coordination and the
Fermi level is roughly at the middle of $d_{3z^2-r^2}$ state
(Fig.~\ref{pt_L1_N1A1}(c)). As a result, $d_{3z^2-r^2}$ orbital is
more occupied than $d_{x^2-y^2}$. This corresponds to a negative
orbital polarization, see Fig.~\ref{pt_L1_N1A1}(d). Again, this
negative orbital polarization at the surface Ni state is a direct
result of its local geometry and present without any additional strain
effects.

Let us now turn our attention to the effects of the strain. We took
the relaxed LNO cubic bulk lattice parameter ($\sim 3.81$~\AA) as a
reference, and applied compressive and tensile strain by setting the
in-plane lattice parameter to be smaller and larger, respectively. The
$\pm 3$\% of strain considered here is in a reasonable range to
simulate the experimental situation. For example, LAO, a widely used
substrate for the compressive strain, has a lattice parameter of
3.78~\AA, and SrTiO$_3$, a popular substrate material for compressive
strain, has 3.91~\AA.

\begin{figure}[t]
  \centering \includegraphics[width=4cm,angle=270]{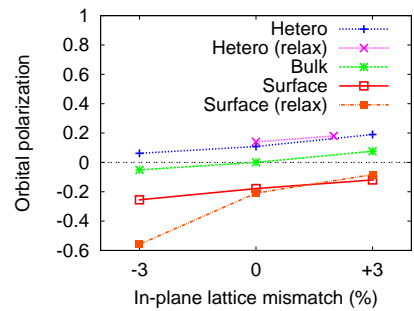}
  \caption{(Color online) Calculated orbital polarizations for a
    LNO/LAO heterostructure (blue), bulk LNO (green), and unrelaxed
    (red) and relaxed slab (orange) as a function of
    strain. Horizontal dotted line shows the zero polarization
    points. The relaxed result for heterostructure (magenta) is from
    Ref.~\cite{MJHan} with re-normalization.
    \label{orbpol}}
\end{figure}

Fig.~\ref{pdos_L1_N1A1}(a) and (b) shows Ni DOS in the heterostructure
under compressive ($-3\%$) and tensile ($+3\%$) strain,
respectively. Similarly, Fig.~\ref{pdos_L1_N1A1}(c) and (d) represents
surface Ni DOS for $-3\%$ and +3$\%$ strain, respectively.  The strain
shifts the relative energy positions of the bands.  However, the
changes in orbital occupation are small and no sign reversal in the
orbital polarization is found. The calculated polarization of the
compressive and tensile strained heterostructure is 0.06
(Fig.~\ref{pdos_L1_N1A1}(a)) and 0.19 (Fig.~\ref{pdos_L1_N1A1}(b)),
respectively, and the polarization of surface is $-0.25$
(Fig.~\ref{pdos_L1_N1A1}(c)) for compressive case and $-0.12$ for
tensile (Fig.~\ref{pdos_L1_N1A1}(d)).

The results of the changes in orbital polarization are summarized as a
function of strain in Fig.~\ref{orbpol}. For bulk LNO, the two $e_g$
orbital states are degenerate under the cubic symmetry, and strain
plays the major role in determining the sign of orbital
polarization. Under the $-3\%$ of compressive strain, $d_{3z^2-r^2}$
orbital is more occupied leading to negative orbital polarization, and
the compressive strain results in the positively polarized orbital
occupation. Therefore the simple electrostatic picture works well in
the bulk case. For heterostructures and surfaces, however, the
symmetry breaking leads those systems to have asymmetric electronic
structures and orbital occupations even without any strain as
discussed in the above. The lattice strain is playing its role on top
of the modified electronic structures, and does not change the sign of
polarization within the range of $\pm 3\%$ lattice mismatch. As is
clearly seen in Fig.~\ref{orbpol}, in the heterostructure, Ni orbitals
are positively polarized and remain positive down to $-3\%$ strain. In
the surface case, the negatively polarized orbital states is retained
up to $+3\%$.

\begin{figure}[t]
  \centering \includegraphics[width=5cm]{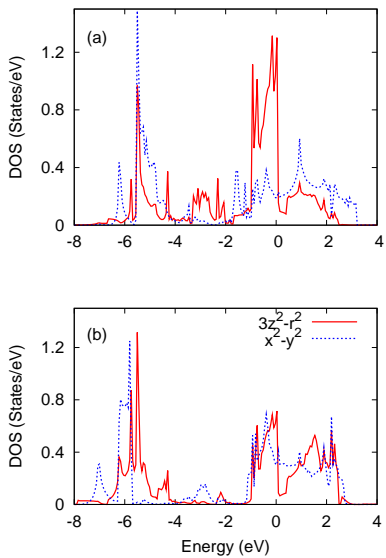}
   \caption{(Color online) Projected Ni-$e_g$ DOS in the
    double layer slab structure. (a) represents to the top layer 
    Ni site that has only five neighboring oxygens, and (b) 
    shows the second layer Ni site that has six neighbors.
    Red (solid) and blue (dashed) lines correspond to $d_{3z^2-r^2}$ 
    and  $d_{x^2-y^2}$ states, respectively, and  Fermi level 
    is set to be zero.
    \label{pdos_L2_both}}
\end{figure}

Rearrangements of atomic positions at the surface often play an
important role and cause a significant change in the electronic
structure. It is therefore important to check the relaxation-induced
change of the electronic structure and orbital occupation. From the
force minimization calculations, it is found that the basic features
of the electronic structure and orbital polarization at the surface do
not change even after the relaxation. We performed the calculations
within the three different in-plane lattice parameters corresponding
to the compressive, zero and tensile strain, and the result of the
calculated orbital polarizations are presented in Fig.~\ref{orbpol}
(dashed-dotted line with filled square boxes) from which one can see
some deviation of the polarization trend but the basic feature is
retained.  Ionic movements make polarization slightly smaller and
larger under the zero and $+3\%$ strain, respectively, with respect to
the unrelaxed results (solid red line with open boxes). It is found
that the relaxation-induced change becomes much larger in the
compressively strained situation, which is attributed to the large
ionic displacement along the out-of-plane direction caused by the
compressive in-plane strain, and the volume conservation often being
observed in the related materials \cite{LNO_film_2,John-arxiv}. In the
zero strain case, the optimized distance between Ni and apical oxygen
is $\sim 2.04$~\AA ~. It is decreased by $4\%$ under the tensile
strain and increased by $11\%$ under the compressive strain.  La ions
also moved slightly in response to strains; toward and out-of the
NiO$_2$ plane by ~0.04 and 0.09 \AA ~ in case of tensile and
compressive strain, respectively. In-plane oxygens (in the NiO$_2$
plane) are found to move toward LaO plane by ~0.06 \AA ~ in case of
compressive strain while it stays at almost same position under
tensile strain. It should be noted that the calculated ionic movements
is from the very thin slab calculation. In realistic situation in
which the film thickness is much larger ($\geq 10~nm$) and it is on
the substrate, the induced polarization change is much smaller and
therefore is closer to the unrelaxed result (Fig.~\ref{orbpol}).

Another important issue regarding the surface is the thickness of
surface-like layers. To understand this point, we performed two-layer
slab calculation. In this structure, the first layer Ni ion ({\it
  i.e.,} Ni(1) in Fig.~\ref{unitcell}) has the reduced coordination,
and the second layer Ni site ({\it i.e.,} Ni(2) in
Fig.~\ref{unitcell}) has the octahedron oxygen environment as in the
bulk. As shown in Fig.~\ref{pdos_L2_both}(b), the cubic symmetry is
significantly restored at the second-layer Ni site and the two $e_g$
states become bulk-like. The difference in the two $e_g$ orbital
occupation is actually small, and the calculated orbital polarization
is $-0.02$. The small deviation between $d_{3z^2-r^2}$ and
$d_{x^2-y^2}$ is the proximity to the surface state. The first layer
Ni is similar to the single layer case (Fig.~\ref{pdos_L2_both}(a)),
the polarization of which is $-0.42$.  In addition, it is noted that
the $d_{3z^2-r^2}$ DOS at the Fermi level is reduced in Ni(2). Large
DOS in Ni(1) as also observed in single-layer case may indicate the
possible unstability of the surface and the other reactions such as
chemisorptions or stoichiometric changes.

Our results shed new light on the interpretation of recent
experiments. It is found from XLD measurement that compressively
strained LNO thin film has negative orbital polarization whereas the
tensile strain does not produce the sign reversal in the polarization
\cite{Jak-arxiv}, which is in a contradiction to the simple
electrostatic Madelung picture assuming the same energy levels both
for $d_{x^2-y^2}$ and $d_{3z^2-r^2}$ ~\cite{Jak-arxiv}. Our
calculation provides a natural explanation as a surface effect
dominant situation; {\it i.e.,} the electronic structure change mainly
caused by the reduced coordination is responsible for the negative
polarization, and the effect persists even under the tensile
strain. The case of heterostructure is more delicate especially
because the XLD shows no orbital polarization in the tensile strain
case for which the simple electrostatic theory predicts the positive
orbital polarization. According to our calculations, the tensile
strain should enhance the positive orbital polarization in
heterostructure. The observation can be understood if there is some
contribution from surface-like states even for the heterostructure
case. As the $d_{x^2-y^2}$ orbital is preferred in the heterostructure
while $d_{3z^2-r^2}$ is at the surface, there can be a compensation
between the two which may result in the negligible polarization along
with the relatively small effect from the strains.

\section{Summary}
In conclusion, using density-functional theory, 
we have shown the change in the electronic structure of nickelates
as a function of reduced dimensionality and strain.
In surface Ni, the reduction of the coordination leads to a significant local 
crystal field. This leads to  a negative orbital polarization
both in the absence and presence of strain effects. It is notably different
from the positive orbital polarization found in heterostructures where the
asymmetric DOS caused by the strongly reduced hybridization  along the out-of-plane
direction results in a higher electron occupation of the $x^2-y^2$
orbitals. Strain is found to play a relatively minor role on top of
the modified electronic structure of each system, and the sign of
polarization does not change under the moderate strains. Our results
are providing new insights for understanding recent experiments on
the related systems. 

\section{Acknowledgments}
We thank Jak Chakhalian, Jian Liu, and John Freeland for useful
discussion. This work was supported by the U.S. Department of Energy
(DOE), DE-FG02-03ER46097, and NIU’s Institute for Nanoscience,
Engineering, and Technology. Work at Argonne National Laboratory was
supported by the U.S.  DOE, Office of Science, Office of Basic Energy
Sciences, under Contract No. DE-AC02-06CH11357.

\end{document}